\documentclass{article}
\usepackage{amsmath}
\usepackage{graphicx}
\usepackage{geometry}
\usepackage{natbib}
\title{Modeling Response Consistency in Multi-Agent LLM Systems: A Comparative Analysis of Shared and Separate Context Approaches}
\author{Tooraj Helmi \\
TUniversity of Southern California \\
\texttt{thelmi@usc.edu}
}

\begin{document}

\maketitle

\begin{abstract}
Large Language Models (LLMs) are increasingly utilized in multi-agent systems (MAS) to enhance collaborative problem-solving and interactive reasoning. Recent advancements have enabled LLMs to function as autonomous agents capable of understanding complex interactions across multiple topics. However, deploying LLMs in MAS introduces challenges related to context management, response consistency, and scalability, especially when agents must operate under memory limitations and handle noisy inputs. While prior research has explored optimizing context sharing and response latency in LLM-driven MAS, these efforts often focus on either fully centralized or decentralized configurations, each with distinct trade-offs.

In this paper, we develop a probabilistic framework to analyze the impact of shared versus separate context configurations on response consistency and response times in LLM-based MAS. We introduce the Response Consistency Index (RCI) as a metric to evaluate the effects of context limitations, noise, and inter-agent dependencies on system performance. Our approach differs from existing research by focusing on the interplay between memory constraints and noise management, providing insights into optimizing scalability and response times in environments with interdependent topics. Through this analysis, we offer a comprehensive understanding of how different configurations impact the efficiency of LLM-driven multi-agent systems, thereby guiding the design of more robust architectures.
\end{abstract}

\section{Introduction}
\label{sec:intro}
Large Language Models (LLMs) such as GPT-3 \citep{brown2020gpt3}, BERT \citep{devlin2018bert}, and T5 \citep{raffel2020exploring} have demonstrated exceptional capabilities in natural language understanding and generation, leading to widespread adoption in applications like chatbots, content generation, and automated support systems. However, deploying LLMs as agents in complex, multi-topic environments presents unique challenges. These challenges primarily revolve around managing limited memory and ensuring response consistency when multiple topics and interactions occur concurrently \citep{wooldridge2009introduction, stone2000multiagent}.

The introduction of multi-agent LLM systems offers a promising solution by distributing tasks across specialized agents, each focusing on specific contexts or topics. Notably, recent research has explored using multi-agent LLMs for collaborative problem-solving and distributed AI \citep{shehory1998methods, stone2019multiagent, tambe1997towards}. However, this approach introduces additional considerations, particularly regarding response times, as agents may need to query each other to access relevant context \citep{kaminka2002multi}. Unlike traditional systems where multiple agents increase noise due to redundant processing, LLM-based multi-agent systems' primary challenge is increased response time due to inter-agent queries rather than a direct increase in noise.

In multi-agent systems (MAS), efficient context management is crucial for maintaining response consistency, especially in dynamic, real-time environments with interdependent topics \citep{ferber1999multiagent, shehory1998methods}. Traditionally, MAS configurations have leveraged centralized databases \citep{bernstein1986concurrency} or distributed grids \citep{foster2001grid} to optimize data management. However, these approaches may become less effective when applied to LLM-based systems where memory limitations and context overflow are significant concerns.

Existing configurations for MAS include fully centralized models where a single agent handles all tasks, decentralized models where agents operate independently, and hybrid models that combine centralized control with distributed autonomy \citep{parunak1999industrial, tambe1997towards}. Centralized systems efficiently coordinate tasks but may struggle with scalability due to context overflow as the number of topics increases. On the other hand, decentralized systems reduce context overflow but increase response time as agents need to query each other for missing context \citep{kaminka2002multi, shehory1998multi}. Hybrid approaches balance shared knowledge with specialized processing, allowing agents to operate semi-autonomously \citep{ross2019probability}.

This paper focuses on two specific configurations for LLM-based agents. The first approach uses a single agent with a shared context, where all topics utilize centralized memory. While this configuration simplifies context management, it suffers from context overflow as the number of topics increases. The second configuration involves multiple agents, each with its own local context, thereby reducing the risk of context overflow. However, the need for inter-agent querying in this configuration can lead to increased response times, especially when context retrieval is necessary.

The key challenge is determining how these configurations impact response consistency and response time in environments where agents operate with limited memory and potentially noisy inputs. To address this, we develop a probabilistic framework to analyze the effects of memory limitations, noise, and inter-topic dependencies on response accuracy. We introduce the Response Consistency Index (RCI) as a metric to quantify the impact of these factors, allowing us to evaluate trade-offs between scalability, response consistency, and performance in both shared and separate context configurations.

\section{Related Work}
\label{sec:related_work}

The integration of Large Language Models (LLMs) into multi-agent systems (MAS) has become a significant area of research in recent years, especially with the advancements in LLM capabilities. Traditionally, MAS research has focused on optimizing task allocation, communication, and coordination among agents \citep{wooldridge2009introduction, shehory1998methods, tambe1997towards}. However, the introduction of LLMs into these systems presents unique challenges and opportunities.

Recent studies have explored various frameworks for utilizing LLMs within MAS, aiming to enhance collaborative problem-solving and interactive reasoning. For example, \citet{qian2024macnet} introduced MacNet, a framework that organizes agents using directed acyclic graphs to facilitate structured communication and collaborative reasoning. The study demonstrated that increasing the number of LLM-based agents can lead to emergent behaviors that improve overall problem-solving efficiency.

The LLMArena framework by \citet{chen2024llmarena} focuses on evaluating LLM capabilities in dynamic multi-agent environments. This framework provides a quantitative assessment of agents' spatial reasoning, strategic planning, and team collaboration. Such evaluations are critical for understanding the limitations of LLMs when applied in MAS setups where context management and response times are key challenges.

Moreover, \citet{xie2024trustsim} explored the use of LLM agents to simulate human trust behaviors in virtual environments. This research contributes to the understanding of social dynamics in multi-agent interactions, highlighting how LLMs can simulate complex human-like behaviors, which is valuable for applications in areas like autonomous negotiation and collaborative simulations.

Another prominent area of recent research focuses on optimizing context management and reducing response latency in LLM-driven MAS. For instance, \citet{zhao2023contextsharing} explored strategies to minimize the overhead associated with inter-agent context queries by implementing efficient memory-sharing techniques. Their findings suggest that while shared memory models can reduce redundancy, they also increase response times due to context overflow as the number of agents increases.

\citet{li2023decentralizedmas} proposed a decentralized architecture where each agent maintains its own local context, reducing the likelihood of context overflow but introducing latency due to inter-agent synchronization. Their work aligns with the observations made by \citet{kaminka2002multi}, who noted that decentralized systems, while scalable, often suffer from increased response times when agents must frequently query each other for missing information.

The challenge of managing noise in LLM-driven multi-agent systems has also been explored. \citet{huang2023noisereduction} introduced techniques for noise filtering in collaborative LLM environments, showing that reducing noise in inter-agent communication improves response consistency. However, as highlighted by \citet{durfee1991distributed} and \citet{foster2001grid}, the overhead of noise reduction can still impact system scalability, particularly in high-load environments.

While many studies focus on optimizing the performance of LLMs in MAS, few have addressed the trade-offs between shared and separate context configurations. For instance, \citet{parunak1999industrial} discussed hybrid approaches where agents use shared contexts for common tasks while retaining local contexts for specialized operations. \citet{ross2019probability} expanded on this by analyzing the impact of context-sharing on scalability and noise management, noting that while shared contexts are efficient, they are prone to overflow with increasing task complexity.

In contrast to prior research that primarily addresses collaborative reasoning and task allocation, our work focuses on the impact of memory limitations and noise on response consistency and response times in LLM-based MAS. By introducing a probabilistic framework and defining the Response Consistency Index (RCI), we aim to provide a deeper understanding of how shared and separate context configurations affect the scalability and performance of these systems under varying conditions.

\section{Approach}
\label{sec:approach}

In this section, we develop a probabilistic framework to evaluate the impact of memory limitations and noise on both \textbf{response consistency} and \textbf{response time} in LLM-based systems. We focus on two configurations: a single agent with a shared context and multiple agents with separate contexts. The framework utilizes Poisson processes and exponential decay to model statement generation, context retention, and noise propagation.

\subsection{Response Consistency Index (RCI)}

The generation of both correct and noisy statements is modeled using Poisson processes. This distribution captures the sporadic arrival of user inputs due to its memoryless property. Let \( \lambda_i^{\text{correct}} \) and \( \lambda_i^{\text{noise}} \) represent the rates at which correct and noisy statements are generated for topic \( i \). The combined rate is defined as:
\[
\lambda_i^{\text{total}} = \lambda_i^{\text{correct}} + \lambda_i^{\text{noise}}
\]
Given the limited memory window \( M \), the probability of retaining a correct statement diminishes as more statements are generated. This probability is modeled using exponential decay:
\[
P(\text{Correct Within Memory}_i) = e^{-\lambda_i^{\text{total}} M}
\]
Here, \( M \) represents the duration for which the system can retain historical context before older information is discarded, measured in units of time corresponding to the rate \( \lambda \).

Noise can disrupt response consistency, especially if a noisy statement follows a correct one. The probability that a noisy statement follows a correct statement is:
\[
P(\text{Noise After Correct}_i) = e^{-\lambda_i^{\text{total}} M} \times \frac{\lambda_i^{\text{noise}}}{\lambda_i^{\text{total}}}
\]
To account for inter-topic dependencies in multi-agent systems, we introduce a correlation matrix \( \rho_{i,j} \):
\[
P(\text{Noise Impact}_i) = P(\text{Noise After Correct}_i) + \sum_{j \neq i} \rho_{i,j} \cdot P(\text{Noise After Correct}_j)
\]

\paragraph{RCI for Shared and Separate Context Models}
In the shared context model, all topics share a single memory window:
\[
\Lambda = \sum_{i} \lambda_i^{\text{total}}
\]
\[
\text{RCI}_{\text{shared}} = \left(1 - e^{-\Lambda M}\right) \times \left[1 - \left(\sum_{i} e^{-\Lambda M} \frac{\lambda_i^{\text{noise}}}{\Lambda} + \sum_{i} \sum_{j \neq i} \rho_{i,j} e^{-\Lambda M} \frac{\lambda_j^{\text{noise}}}{\Lambda}\right)\right]
\]

In the separate context model:
\[
\text{RCI}_{\text{separate}} = \prod_{i} \left[\left(1 - e^{-\lambda_i^{\text{total}} M_i}\right) \times \left(1 - \left(e^{-\lambda_i^{\text{total}} M_i} \frac{\lambda_i^{\text{noise}}}{\lambda_i^{\text{total}}} + \sum_{j \neq i} \rho_{i,j} e^{-\lambda_j^{\text{total}} M_j} \frac{\lambda_j^{\text{noise}}}{\lambda_j^{\text{total}}}\right)\right)\right]
\]

\paragraph{RCI Ratio}
To compare the performance of the two models, we define the \textbf{RCI Ratio}:
\[
\text{RCI Ratio} = \frac{\text{RCI}_{\text{separate}}}{\text{RCI}_{\text{shared}}}
\]
A ratio greater than 1 indicates that the separate context model is more consistent, while a ratio less than 1 indicates better performance of the shared context model.

\subsection{Response Time Analysis}

In addition to consistency, another critical metric is the \textbf{response time}. The response time is influenced by the time to search within the memory window and, for separate contexts, the time needed to query context from other agents.

The search time within the memory window is modeled as:
\[
T_{\text{search}}(M) = \alpha \log(1 + M)
\]
For separate contexts, querying other agents introduces additional overhead:
\[
T_{\text{query}}(N) = \beta N
\]

\paragraph{Response Time for Shared and Separate Contexts}
For the shared context model:
\[
T_{\text{shared}} = \alpha \log(1 + M)
\]
For the separate context model:
\[
T_{\text{separate}} = \alpha \log(1 + M_{\text{separate}}) + \beta N
\]

\paragraph{Response Time Ratio}
To compare the efficiency of the two models, we define the \textbf{Response Time Ratio}:
\[
\text{Response Time Ratio} = \frac{T_{\text{separate}}}{T_{\text{shared}}} = 1+\frac{\beta N}{\alpha \log(1 + M)}
\]

If the ratio is greater than 1, the separate context model incurs additional overhead, making it slower. Conversely, if the ratio is less than 1, the separate context model is faster due to reduced search time in isolated contexts.

By analyzing both RCI and response time, we can evaluate the trade-offs between scalability, noise tolerance, and performance in shared versus separate context configurations.

\subsection{Model Implications}

In this subsection, we simplify the derived formulas to better understand the implications of our models on system performance and visualize the impact of key parameters on both the Response Consistency Index (RCI) and response time.

To gain insights into how different configurations impact the RCI and response time, we consider certain simplifications. For instance, assuming equal rates for both topics (\( \lambda_1^{\text{total}} = \lambda_2^{\text{total}} = \lambda^{\text{total}} \)) and symmetric correlations (\( \rho_{1,2} = \rho_{2,1} = \rho \)), the RCI for the shared and separate context models can be simplified as follows:

\[
\text{RCI}_{\text{shared}} = \left(1 - e^{-2 \lambda^{\text{total}} M}\right) \times \left(1 - e^{-2 \lambda^{\text{total}} M} \left(1 + \frac{\rho}{2}\right) \frac{\lambda^{\text{noise}}}{\lambda^{\text{total}}}\right)
\]

\[
\text{RCI}_{\text{separate}} = \left(1 - e^{-\lambda^{\text{total}} M}\right)^2 \times \left(1 - \left(e^{-\lambda^{\text{total}} M} + \rho e^{-\lambda^{\text{total}} M}\right) \frac{\lambda^{\text{noise}}}{\lambda^{\text{total}}}\right)^2
\]

To compare the two configurations, we derive the simplified RCI Ratio:
\[
\text{RCI Ratio} = \frac{\text{RCI}_{\text{separate}}}{\text{RCI}_{\text{shared}}} = \frac{\left(1 - e^{-\lambda^{\text{total}} M}\right)^2 \times \left(1 - \left(e^{-\lambda^{\text{total}} M} + \rho e^{-\lambda^{\text{total}} M}\right) \frac{\lambda^{\text{noise}}}{\lambda^{\text{total}}}\right)^2}{\left(1 - e^{-2 \lambda^{\text{total}} M}\right) \times \left(1 - e^{-2 \lambda^{\text{total}} M} \left(1 + \frac{\rho}{2}\right) \frac{\lambda^{\text{noise}}}{\lambda^{\text{total}}}\right)}
\]

The response time for shared and separate contexts is simplified as:
\[
T_{\text{shared}} = \alpha \log(1 + M)
\]
\[
T_{\text{separate}} = \alpha \log(1 + M) + \beta N
\]
The simplified Response Time Ratio is given by:
\[
\text{Response Time Ratio} = \frac{T_{\text{separate}}}{T_{\text{shared}}} = \frac{\alpha \log(1 + M) + \beta N}{\alpha \log(1 + M)}
\]

Let's explore the implications of the derived models using illustrative graphs to demonstrate how different parameters impact the Response Consistency Index (RCI) for both the single-agent (shared context) and multi-agent (separate context) configurations. Each graph includes two curves: one for the single-agent configuration and one for the multi-agent setup, allowing for direct comparison.

\begin{figure}[h]
    \centering
    \begin{minipage}{0.49\textwidth}
        \includegraphics[width=\textwidth]{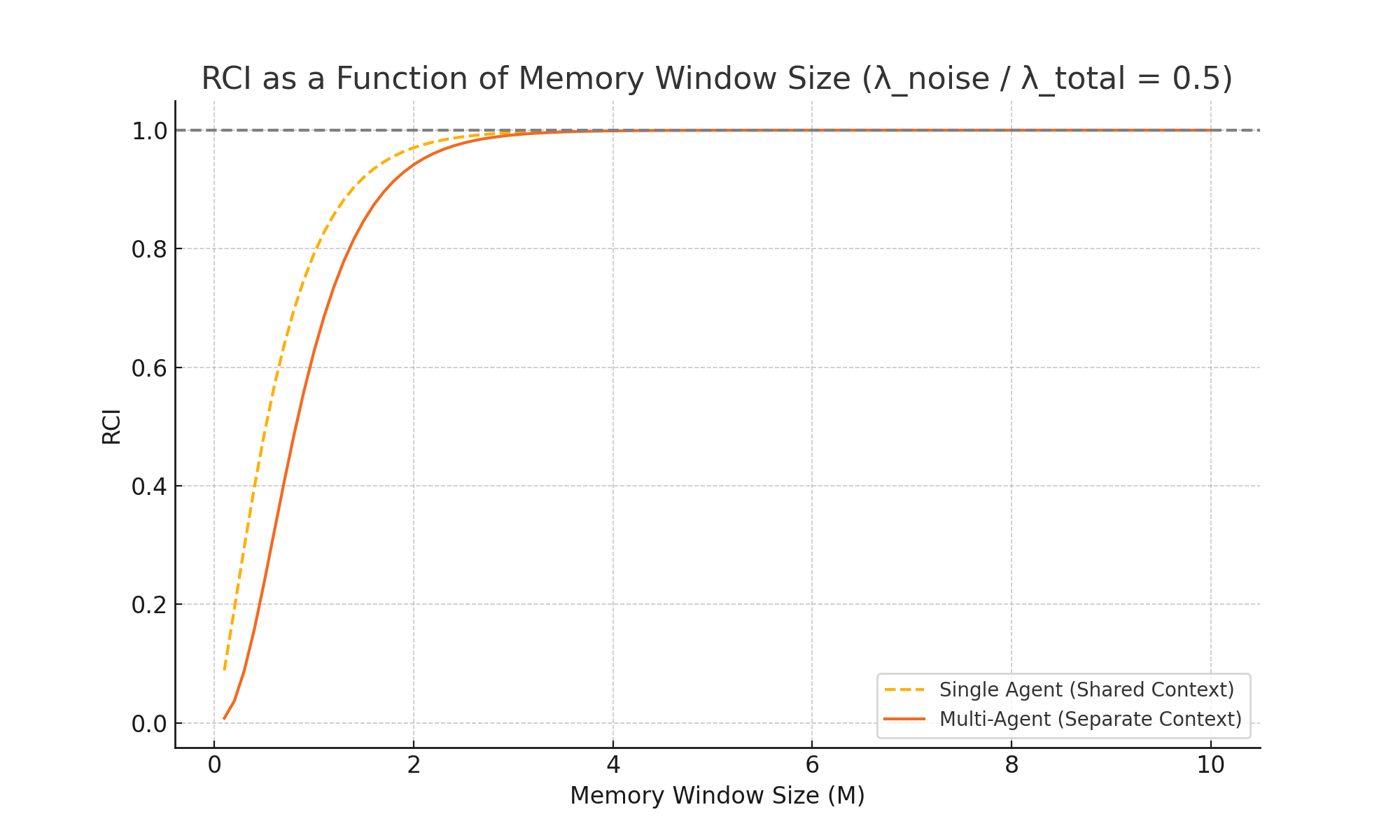}
        \caption{RCI as a function of memory window size (\( \lambda_{\text{noise}} / \lambda_{\text{total}} = 0.5 \)).}
        \label{fig:rci_varying_M}
    \end{minipage}
    \hfill
    \begin{minipage}{0.49\textwidth}
        \includegraphics[width=\textwidth]{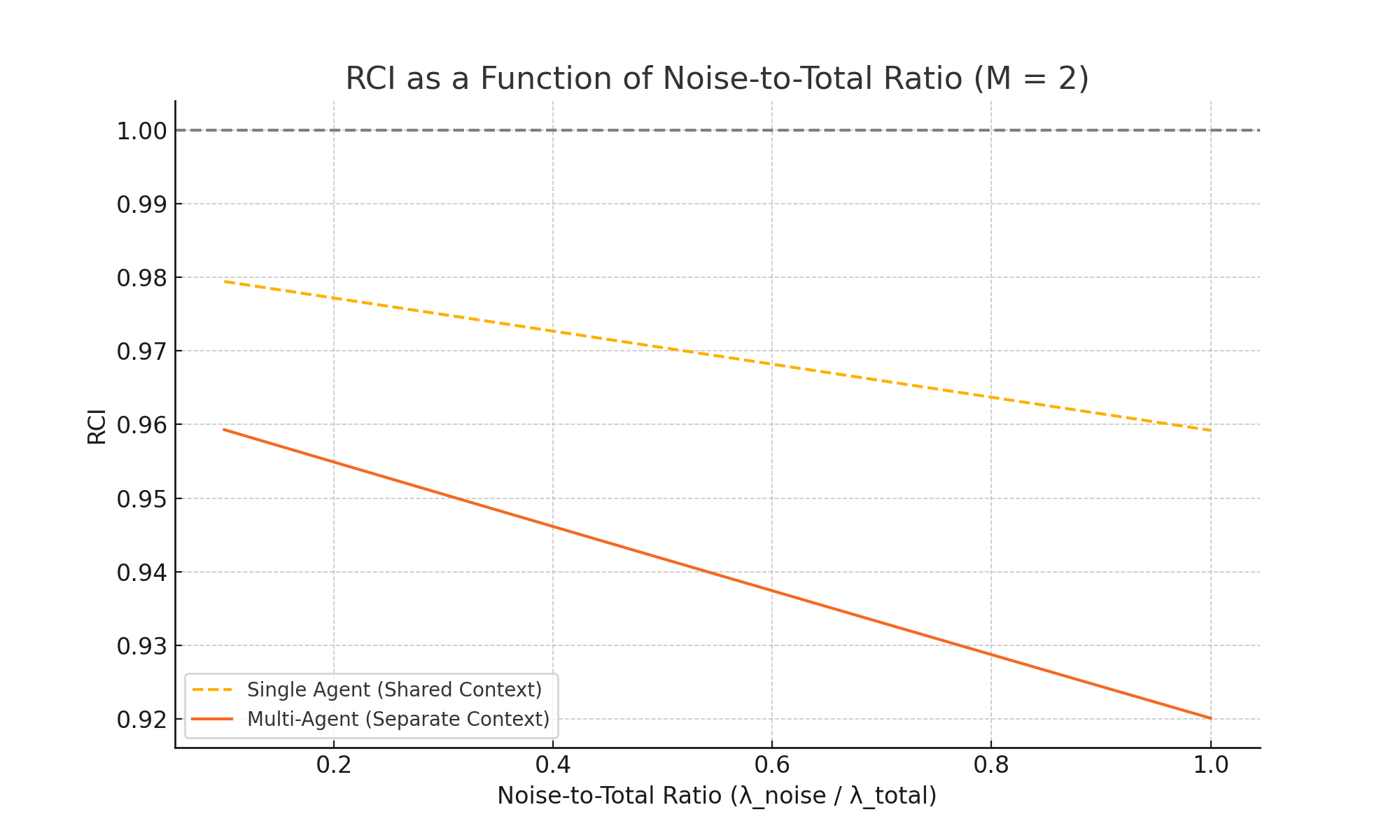}
        \caption{RCI as a function of noise-to-total ratio (\( M = 2 \)).}
        \label{fig:rci_varying_noise}
    \end{minipage}
\end{figure}

Figure~\ref{fig:rci_varying_M} shows how the RCI changes with increasing memory window size (\( M \)), with a fixed noise-to-total ratio of \( \lambda_{\text{noise}} / \lambda_{\text{total}} = 0.5 \). As the memory window size (\( M \)) increases, the response consistency improves for both configurations. This is because larger memory windows allow more context to be retained, reducing the likelihood of losing relevant information. However, the single-agent model, which utilizes a shared context, benefits more significantly from increased memory. In contrast, the multi-agent model suffers from limitations due to its isolated memory windows. This leads to a squared degradation rate, making it less effective as the number of agents increases.

Next, we analyze how the RCI changes as the noise-to-total ratio (\( \lambda_{\text{noise}} / \lambda_{\text{total}} \)) increases, with a fixed memory window size of \( M = 2 \). Figure~\ref{fig:rci_varying_noise} illustrates that increasing the noise ratio results in a significant drop in RCI for both configurations. Higher noise levels introduce incorrect or irrelevant statements, thereby reducing response consistency. The single-agent model is better equipped to handle noise due to its access to a shared context, which allows it to mitigate noise more effectively. Conversely, the multi-agent model is more sensitive to noise due to its separate contexts, where noise introduced in one agent's context cannot be easily corrected by another agent. This results in a more pronounced decline in RCI for the multi-agent system, reflecting its vulnerability to noise.

From the analysis, we observe that:
- For smaller memory windows (\( M \)), the impact of noise is more pronounced, especially for the multi-agent setup, due to limited context retention.
- The single-agent configuration demonstrates better resilience to noise, particularly when \( M \) is increased, leveraging its unified context to counteract the effects of noise.
- As the number of topics increases, the multi-agent setup struggles due to the compounded effect of noise across independent contexts, leading to faster degradation in RCI.

This analysis highlights the trade-offs between using a single shared context versus multiple separate contexts, emphasizing the need to balance memory retention and noise management in multi-agent LLM systems.


\begin{thebibliography}{22}
\providecommand{\natexlab}[1]{#1}
\providecommand{\url}[1]{\texttt{#1}}
\expandafter\ifx\csname urlstyle\endcsname\relax
  \providecommand{\doi}[1]{doi: #1}\else
  \providecommand{\doi}{doi: \begingroup \urlstyle{rm}\Url}\fi

\bibitem[Bernstein and Goodman(1986)]{bernstein1986concurrency}
Philip~A. Bernstein and Nathan Goodman.
\newblock Concurrency control in distributed database systems.
\newblock \emph{ACM Computing Surveys}, 13\penalty0 (2):\penalty0 185--221, 1986.

\bibitem[Chen and Wang(2024)]{chen2024llmarena}
Hong Chen and Jie Wang.
\newblock Llmarena: Evaluating llms in dynamic multi-agent environments.
\newblock \emph{arXiv preprint arXiv:2402.16499}, 2024.

\bibitem[Devlin(2018)]{devlin2018bert}
Jacob Devlin.
\newblock Bert: Pre-training of deep bidirectional transformers for language understanding.
\newblock \emph{arXiv preprint arXiv:1810.04805}, 2018.

\bibitem[Durfee(1991)]{durfee1991distributed}
Edmund~H. Durfee.
\newblock Distributed problem solving and planning.
\newblock \emph{Multiagent Systems}, 3:\penalty0 33--80, 1991.

\bibitem[et~al.(2020{\natexlab{a}})]{raffel2020exploring}
Colin~Raffel et~al.
\newblock Exploring the limits of transfer learning with a unified text-to-text transformer.
\newblock \emph{Journal of Machine Learning Research}, 21\penalty0 (140):\penalty0 1--67, 2020{\natexlab{a}}.

\bibitem[et~al.(2020{\natexlab{b}})]{brown2020gpt3}
Tom B.~Brown et~al.
\newblock Language models are few-shot learners.
\newblock \emph{arXiv preprint arXiv:2005.14165}, 2020{\natexlab{b}}.

\bibitem[Ferber(1999)]{ferber1999multiagent}
Jacques Ferber.
\newblock \emph{Multi-Agent Systems: An Introduction to Distributed Artificial Intelligence}.
\newblock Addison-Wesley, 1999.

\bibitem[Foster and Kesselman(2001)]{foster2001grid}
Ian Foster and Carl Kesselman.
\newblock \emph{The Grid: Blueprint for a New Computing Infrastructure}.
\newblock Morgan Kaufmann, 2001.

\bibitem[Huang and Chen(2023)]{huang2023noisereduction}
Lin Huang and Qiang Chen.
\newblock Noise reduction techniques in collaborative llm environments.
\newblock \emph{arXiv preprint arXiv:2311.09876}, 2023.

\bibitem[Kaminka and Tambe(2002)]{kaminka2002multi}
Gal~A. Kaminka and Milind Tambe.
\newblock Multi-agent teamwork in dynamic and uncertain environments.
\newblock \emph{Artificial Intelligence}, 142:\penalty0 1--5, 2002.

\bibitem[Li and Zhang(2023)]{li2023decentralizedmas}
Xing Li and Hao Zhang.
\newblock Decentralized context management in llm-driven multi-agent systems.
\newblock \emph{Artificial Intelligence}, 311:\penalty0 98--115, 2023.

\bibitem[Parunak(1999)]{parunak1999industrial}
H.~Van~Dyke Parunak.
\newblock \emph{Industrial and Practical Applications of DAI}.
\newblock Springer, 1999.

\bibitem[Peter~Stone(2019)]{stone2019multiagent}
Kagan~Tumer Peter~Stone, G. Michael~Young.
\newblock Multiagent learning is not the same as machine learning: A survey.
\newblock \emph{AI Magazine}, 38:\penalty0 17--34, 2019.

\bibitem[Qian and Zhou(2024)]{qian2024macnet}
Li~Qian and Wei Zhou.
\newblock Macnet: Structured communication framework for llm-based multi-agent reasoning.
\newblock \emph{arXiv preprint arXiv:2406.07155}, 2024.

\bibitem[Ross(2019)]{ross2019probability}
Sheldon~M. Ross.
\newblock Introduction to probability models.
\newblock In \emph{Academic Press}, 2019.

\bibitem[Shehory and Kraus(1998{\natexlab{a}})]{shehory1998methods}
Onn Shehory and Sarit Kraus.
\newblock Methods for task allocation via agent coalition formation.
\newblock \emph{Artificial Intelligence}, 101:\penalty0 165--200, 1998{\natexlab{a}}.

\bibitem[Shehory and Kraus(1998{\natexlab{b}})]{shehory1998multi}
Onn Shehory and Sarit Kraus.
\newblock Task allocation in mas using combinatorial auctions.
\newblock In \emph{AAAI Conference on Artificial Intelligence}, pages 167--172, 1998{\natexlab{b}}.

\bibitem[Stone and Veloso(2000)]{stone2000multiagent}
Peter Stone and Manuela Veloso.
\newblock Multiagent systems: A survey from a machine learning perspective.
\newblock In \emph{Autonomous Robots}, volume~8, pages 345--383, 2000.

\bibitem[Tambe(1997)]{tambe1997towards}
Milind Tambe.
\newblock Towards flexible teamwork.
\newblock \emph{Journal of Artificial Intelligence Research}, 7:\penalty0 83--124, 1997.

\bibitem[Wooldridge(2009)]{wooldridge2009introduction}
Michael Wooldridge.
\newblock \emph{An Introduction to MultiAgent Systems}.
\newblock Wiley, 2009.

\bibitem[Xie and Liu(2024)]{xie2024trustsim}
Dong Xie and Zhen Liu.
\newblock Trustsim: Simulating human trust in multi-agent systems using llms.
\newblock \emph{Journal of Artificial Intelligence Research}, 75:\penalty0 25--48, 2024.

\bibitem[Zhao and Li(2023)]{zhao2023contextsharing}
Mei Zhao and Yu~Li.
\newblock Optimizing context sharing in multi-agent systems using large language models.
\newblock \emph{IEEE Transactions on Neural Networks and Learning Systems}, 2023.

\end{thebibliography}
\end{document}